\documentstyle{article}

\def\abstract#1{\vskip 7mm 
        \begin{center}{\large Abstract}\par \smallskip
                \begin{minipage}[c]{12cm}
                        \small #1
                \end{minipage}
        \end{center}
}
\def\title#1{\begin{center}{\Large\bf #1}\end{center}}
\def\author#1{\vskip 5mm \begin{center}{#1}\end{center}}
\def\address#1{\begin{center}{\it #1}\end{center}}

\newtheorem{theorem}     {Theorem}
\newtheorem{lemma}       {Lemma}

\newtheorem{conjecture}  {Conjecture}

\begin{document}

\title{%
On the existence of global solutions for $T^{3}$-Gowdy spacetimes 
with stringy matter
}
\author{%
Makoto Narita\footnote{Present address:
Max-Planck-Institut f\"ur Gravitationsphysik, 
Albert-Einstein-Institut, 
Am M\"uhlenberg 1, 
14476 Golm,
Germany,
E-mail: maknar@aei-potsdam.mpg.de}}
\address{%
  Advanced Research Institute for Science and Engineering, Astrophysics group, 
  Waseda University,
  Shinjuku, Tokyo 169-8555, Japan, 
E-mail: narita@gravity.phys.waseda.ac.jp
}
\abstract{
We show a global existence theorem for Einstein-matter equations of 
$T^{3}$-Gowdy symmetric spacetimes with stringy matter.  
The areal time coordinate is used.  
It is shown that this spacetime has a crushing singularity 
into the past.  
From these results we can show that 
the spacetime is foliated by compact hypersurfaces of 
constant mean curvature.  
}
PACS: $02.03.J_{r}, 04.20.D_{W}, 04.20.E_{X}, 98.80.H_{W}$

\newpage

\section{Introduction}
\label{intro}

The singularity theorem states that generic spacetimes have 
a singularity (spacetimes with incomplete geodesics).  
If the singularity can be seen, i.e. there exists a {\it naked singularity}, 
the predictability is violated.  
Thus, the strong cosmic censorship conjecture proposed by Penrose is 
the most important (and an unsolved) problem 
in classical general relativity.  
Roughly speaking, the conjecture states 
that naked singularities should not evolve from regular initial data. 
However, this statement remains elusive.  
Then, a sharp formulation that seems provable is given as follows: 
%%%%%%%%%%%%%%%%%%%%%%%%%%%%%%%%%%%%%%%%%%%%%%%%%%%%%%%%%%%%%%%%%%%%%%%%%%
\begin{conjecture}~\cite{AL99,CD,KN,ME}
\label{Cj-AL}
Let $M$ be a partial Cauchy surface and $\Phi_{i}$ are matter fields on $M$.  
Then for generic data sets $(M, h, k, \Phi_{i})$, 
where $h$ and $k$ are the first and 
second fundamental forms of $M$, the maximal Cauchy development of 
$(M, h, k, \Phi_{i})$ is equal to the maximal extension of 
$(M, h, k, \Phi_{i})$.  
\hfill$\Box$
\end{conjecture}
%%%%%%%%%%%%%%%%%%%%%%%%%%%%%%%%%%%%%%%%%%%%%%%%%%%%%%%%%%%%%%%%%%%%%%%%%%%
To answer this problem we must prove (1) a global existence theorem for the 
Einstein-matter equations with a {\it suitable} time coordinate and (2) 
inextendibility of the maximal Cauchy development of 
initial data.  This paper is mainly related to the above problem (1). 
It is very difficult to answer the conjecture in general by using present 
mathematical techniques.  
Then, it is necessary to make some simplifications, which are 
symmetry assumptions or restriction on initial data.

For the case of small initial data without symmetry assumptions, 
partial results 
for this problem are provided by 
Christodoulou-Klainerman theorem which states that 
any asymptotically flat initial data set which is sufficiently 
close to the trivial one has a complete maximal future development
~\cite{CK,KN}.  
Recently, Andersson and Moncrief have shown that 
for a vacuum data set which is sufficiently close to the data 
in a spatially compact (local) open Friedmann-Robertson-Walker 
spacetime, the maximal Cauchy 
development is causally geodesically complete in the expanding 
direction~\cite{AL99} (see also Ref.~\cite{R02}).

For the case of large initial data with symmetry assumptions, 
some results have been obtained.  
(Hereafter, models we consider are restricted to spatially compact 
spacetimes.) 
The first result for inhomogeneous cosmological spacetimes was proved by 
Moncrief~\cite{M81}. 
He proved a global existence theorem for $T^{3}$-Gowdy spacetimes 
(which are vacuum spacetimes with $U(1)\times U(1)$ symmetry and 
whose spatial topology is $T^{3}$) in the areal coordinate 
(defined in section~\ref{Gowdy-EMDA}).  
This result has been generalized to the case with non-vanishing 
twist~\cite{BCIM} or with Vlasov matter~\cite{AH}.  
A result on global existence in the areal time 
has also been obtained for hyperbolic symmetric spacetimes with 
Vlasov matter~\cite{ARR}.

Although the areal time coordinate is well-chosen one in the sense 
that it is geometrically defined, this coordinate 
choice strongly depends on spacetime symmetry.  
The most attractive (and independent on spacetime symmetry) 
time coordinate is the constant mean curvature (CMC) 
one. 
CMC foliations can avoid a crushing singularity which is one 
where there is a foliation on a neighborhood of the singularity 
whose mean curvature tends uniformly to infinity as the singularity is 
approaches~\cite{ES}. 
It was shown from Hawking's singularity theorem 
that the  crushing singularity is either a true singularity 
or a boundary of maximal Cauchy development, i.e., a Cauchy horizon. 
Thus, the existence of CMC foliations is closely related to 
conjecture~\ref{Cj-AL}. 
Indeed, it has been conjectured as follows:
%%%%%%%%%%%%%%%%%%%%%%%%%%%%%%%%%%%%%%%%%%%%%%%%%%%%%%%%%%%%%%%%%%%%%%%%%%%
\begin{conjecture}~\cite{ME}
\label{Cj-ME}
Every maximally extended, globally hyperbolic 
spacetime can be foliated by CMC hypersurfaces.
\hfill$\Box$
\end{conjecture}
%%%%%%%%%%%%%%%%%%%%%%%%%%%%%%%%%%%%%%%%%%%%%%%%%%%%%%%%%%%%%%%%%%%%%%%%%%%%
A global existence theorem in the CMC time coordinate for $T^{3}$-Gowdy 
spacetimes was proved in Ref.~\cite{IM}. 
Recently, this result has been generalized to the case of local 
$U(1)\times U(1)$ symmetric spacetimes with Vlasov or wave-map 
matter~\cite{R97,H02b}.  
Results on global existence in the CMC time have been shown for 
hyperbolic symmetric spacetimes with Maxwell field or 
Vlasov matter~\cite{H02a,ARR}.

Note that the choice of matter models is serious.  
For some matter models (Vlasov matter, Maxwell field and wave map), 
there are global existence theorems as above.  
Contrary, a global non-existence result have been obtained 
for the Einstein-dust system~\cite{IR}.  
Therefore, it would be worth to investigate global existence problems for 
several matter models, in particular, systems of 
{\it nonlinear}  and {\it fundamental fields} equations.

The purpose of the present paper is generalization of the above results, 
that is, to show global existence theorems 
for the $T^{3}$-Gowdy symmetric 
spacetimes with stringy matter fields.  
From the unified theoretical point of view, 
there are many reasons to believe that the distinction between 
fundamental fields (i. e. gravitational and matter fields) 
interactions is impossible in 
asymptotic regions (e. g. near singularities) of spacetimes and 
the most consistent theory along these line is superstring/M-theory.  
Therefore, the matter fields we will consider are Maxwell-dilaton-axion 
fields which arise naturally from low-energy effective 
superstring theory~\cite{ON}.  
This system is nonlinear even if 
the background spacetime is flat since there is a dilaton coupling. 
One of the main results of this paper is a global existence theorem 
for such system in the areal time coordinate (Theorem~\ref{main-Th}) 
and another of them is one in the CMC time coordinate 
(Theorem~\ref{main-Th2}).

In section~\ref{Gowdy-EMDA}, we will review 
$T^{3}$-Gowdy symmetric spacetimes in 
the Einstein-Maxwell-dilaton-axion (EMDA) system 
and derive the Einstein-matter equations.  
In section~\ref{global}, we shall show a global existence theorem 
for the system in the areal time coordinate. 
In section~\ref{CMC}, 
it is shown that a global existence theorem 
for the system in the CMC time coordinate. 
In section~\ref{dis}, we discuss on inextendibility of the spacetime. 
%%%%%%%%%%%%%%%%%%%%%%%%%%%%%%%%%%%%%%%%%%%%%%%%%%%%%%%%%%%%%%%%%%%%%%%
\section{$T^{3}$-Gowdy symmetric spacetimes in 
the Einstein-Maxwell-dilaton-axion system}
\label{Gowdy-EMDA}
The action of the EMDA theory ~\cite{STW} is
%%%%%%%%%%%%%%%%%%%%%%%%%%%%%%%%%%%%%%%%%%%%%%%%%%%%%%%%%%%%%%%%%%%%%%%%
\begin{eqnarray}
\label{action}
S=S_{G}+S_{M},
\end{eqnarray}
\begin{eqnarray}
S_{G}=\int d^{4}x\sqrt{-g}\left[-\;^{(4)}\!R\right],
\end{eqnarray}
\begin{eqnarray}
S_{M}:=\int d^{4}x\sqrt{-g}\left[e^{-2a_{M}\phi}F^{2}
+2\left(\nabla\phi\right)^{2}
+\frac{1}{3}e^{-4a_{A}\phi}H^{2}\right]
=\int d^{4}x{\cal L}_{M}, 
\end{eqnarray}
%%%%%%%%%%%%%%%%%%%%%%%%%%%%%%%%%%%%%%%%%%%%%%%%%%%%%%%%%%%%%%%%%%%%%%%%%%%
where $g$ is the determinant of a 4-dimensional spacetime metric $g_{ab}$, 
$\;^{(4)}\!R$ is the Ricci scalar for $g_{ab}$, $F$ is the Maxwell field, 
$\phi$ is the dilaton field,
$H=dB\equiv -\frac{1}{2}e^{4a\phi}*d\kappa$ is the three-index
antisymmetric tensor field dual to the axion field $\kappa$, and
$a_{M}$ and $a_{A}$ are coupling constants.
For simplicity, the Chern-Simon term is neglected.  
Varying the action (\ref{action}) with respect to the functions, 
we get the Einstein-matter equations.  

The metric of Gowdy symmetric spacetimes~\cite{GR} is given by 
%%%%%%%%%%%%%%%%%%%%%%%%%%%%%%%%%%%%%%%%%%%%%%%%%%%%%%%%%%%%%%%%%%%%%%%
\begin{equation}
\label{gowdy-metric}
ds^{2}=e^{\lambda(t,\theta)/2}t^{-1/2}(-dt^{2}+d\theta ^{2})+
R(t,\theta)\left[e^{-Z(t,\theta)}\left(dy+X(t,\theta)dz\right)^{2}
+e^{Z(t,\theta)}dz^{2}\right].
\end{equation}
%%%%%%%%%%%%%%%%%%%%%%%%%%%%%%%%%%%%%%%%%%%%%%%%%%%%%%%%%%%%%%%%%%%%%%%%%
Gowdy symmetric spacetimes have two twist free spacelike Killing vectors
$\partial/\partial y$ and $\partial/\partial z$. 
Properties of the metric (\ref{gowdy-metric}) depend on whether 
$\nabla R$ is timelike, spacelike or null.  
When the metric~(\ref{gowdy-metric}) describes a cosmological 
model, i.e. $\nabla R$ is globally 
timelike and the spatial topology is $T^{3}$ (periodic in $\theta$), 
one can take the function $R(t,\theta)=t$ without loss of generality 
by Gowdy's corner theorem if 
the spacetime is vacuum~\cite{GR,M81}.  
This fact be seen from Einstein's equations, 
\begin{equation}
\label{G-G=0}
G_{tt}-G_{\theta\theta}=\ddot{R}-R''=0,
\end{equation}
where dot and prime denote $t$ and $\theta$ derivatives, respectively.  
In the EMDA system, equation~(\ref{G-G=0}) is not satisfied generically.
However, in the case that the Maxwell field
strength $F_{\mu\nu}=\partial_{\mu}A_{\nu}-\partial_{\nu}A_{\mu}$
has only the following components~\cite{MP}, 
%%%%%%%%%%%%%%%%%%%%%%%%%%%%%%%%%%%%%%%%%%%%%%%%%%%%%%%%%%%%%%%%%%%%%%%
$$
F_{ty}=\dot{\omega}(t,\theta), \;\;\;
F_{\theta y}=\omega '(t,\theta), \;\;\;
F_{tz}=\dot{\chi}(t,\theta), \;\;\;
F_{\theta z}=\chi '(t,\theta),
$$
%%%%%%%%%%%%%%%%%%%%%%%%%%%%%%%%%%%%%%%%%%%%%%%%%%%%%%%%%%%%%%%%%%%%%%%%%%%
and $\phi=\phi(t,\theta)$ and $\kappa=\kappa(t,\theta)$, 
equation~(\ref{G-G=0}) is satisfied.
As similar as the $T^{3}$-vacuum case, 
one can take the function $R(t,\theta)=t$ without loss of generality. 
In this gauge choice, 
the spacetimes have spacelike singularities at $t=0$.  
We call this gauge the {\it areal time coordinate} since $R$ is proportional 
to the geometric area function of the orbit of the isometry group.   

We should mention on another important choice for the Maxwell field. 
In~\cite{WIB}, it was taken that 
$\omega$ and $\chi$ are zero but one of the other 
components of the Maxwell field is non-zero. 
This spacetime is called ``magnetic Gowdy spacetime'' and 
concerns the complementary case with ours.

In this coordinate we obtain the Einstein-matter equations as follows:
%%%%%%%%%%%%%%%%%%%%%%%%%%%%%%%%%%%%%%%%%%%%%%%%%%%%%%%%%%%%%%%%%%%%%%
\begin{eqnarray}
\label{ce1}
\dot{\lambda}-t[\dot{Z}^{2}+Z'^{2}+e^{-2Z}(\dot{X}^{2}+X'^{2})]=4tT_{tt},
\end{eqnarray}
%%%%%%%%%%%%%%%%%%%%%%%%%%%%%%%%%%%%%%%%%%%%%%%%%%%%%%%%%%%%%%%%%%%%%%%
where 
%%%%%%%%%%%%%%%%%%%%%%%%%%%%%%%%%%%%%%%%%%%%%%%%%%%%%%%%%%%%%%%%%%%%%%%
\begin{eqnarray}
\label{ttt}
T_{tt}&=&
(\dot{\phi}^{2}+\phi'^{2})+
\frac{1}{4}e^{4a_{A}\phi}(\dot{\kappa}^{2}+\kappa'^{2}) \\ \nonumber
&+&\frac{1}{t}e^{-2a_{M}\phi}[e^{-Z}\{(X\dot{\omega}-\dot{\chi})^{2}+
(X\omega'-\chi')^{2}\}+e^{Z}(\dot{\omega}^{2}+\omega'^{2})] \\ \nonumber
&=&T_{\theta\theta},
\end{eqnarray}
%%%%%%%%%%%%%%%%%%%%%%%%%%%%%%%%%%%%%%%%%%%%%%%%%%%%%%%%%%%%%%%%%%%%%%%
\begin{eqnarray}
\label{ce2}
\lambda '-2t[\dot{Z}Z'+e^{-2Z}\dot{X}X']=8tT_{t\theta},
\end{eqnarray}
%%%%%%%%%%%%%%%%%%%%%%%%%%%%%%%%%%%%%%%%%%%%%%%%%%%%%%%%
where
%%%%%%%%%%%%%%%%%%%%%%%%%%%%%%%%%%%%%%%%%%%%%%%%%%%%%%%
\begin{eqnarray}
\label{tttheta}
T_{t\theta}=\dot{\phi}\phi'
+\frac{1}{4}e^{4a_{A}\phi}\dot{\kappa}\kappa' +
\frac{1}{t}e^{-2a_{M}\phi}
\{e^{-Z}(X\dot{\omega}-\dot{\chi})(X\omega'-\chi')
+e^{Z}\dot{\omega}\omega'\},
\end{eqnarray}
%%%%%%%%%%%%%%%%%%%%%%%%%%%%%%%%%%%%%%%%%%%%%%%%%%%%%%%%%%%%%%%%%%%%%%
%%%%%%%%%%%%%%%%%%%%%%%%%%%%%%%%%%%%%%%%%%%%%%%%%%%%%%%%%%%%%%%%%%%%%%
\begin{eqnarray}
\label{ee-z}
t^{2}\ddot{Z}+t\dot{Z}-t^{2}Z''&+&t^{2}e^{-2Z}(\dot{X}^{2}-X'^{2}) \nonumber \\
&-&2te^{-2a_{M}\phi}
[-e^{-Z}\{(X\dot{\omega}-\dot{\chi})^{2}-
(X\omega'-\chi')^{2}\}+e^{Z}(\dot{\omega}^{2}-\omega'^{2})
]=0,
\end{eqnarray}
%%%%%%%%%%%%%%%%%%%%%%%%%%%%%%%%%%%%%%%%%%%%%%%%%%%%%%%%%%%%%%%%%%%%%%%%%
\begin{eqnarray}
\label{ee-x}
t^{2}\ddot{X}+t\dot{X}-t^{2}X''-2t^{2}(\dot{X}\dot{Z}-X'Z')
-4te^{-2a_{M}\phi}e^{Z}[(\dot{\omega}^{2}-\omega'^{2})X
-(\dot{\omega}\dot{\chi}-\omega'\chi')]=0,
\end{eqnarray}
%%%%%%%%%%%%%%%%%%%%%%%%%%%%%%%%%%%%%%%%%%%%%%%%%%%%%%%%%%%%%%%%%%%%%%%%%
\begin{eqnarray}
\label{ee-phi}
t^{2}\ddot{\phi}+t\dot{\phi}-t^{2}\phi ''
&-&\frac{a_{A}}{2}t^{2}e^{4a_{A}\phi}(\dot{\kappa}^{2}-\kappa '^{2}) 
\nonumber \\
&+&a_{M}te^{-2a_{M}\phi}
[e^{-Z}\{(X\dot{\omega}-\dot{\chi})^{2}-
(X\omega'-\chi')^{2}\}+e^{Z}(\dot{\omega}^{2}-\omega'^{2})
]=0,
\end{eqnarray}
%%%%%%%%%%%%%%%%%%%%%%%%%%%%%%%%%%%%%%%%%%%%%%%%%%%%%%%%%%%%%%%%%%%%%%%%
\begin{eqnarray}
\label{ee-kappa}
t^{2}\ddot{\kappa}+t\dot{\kappa}-t^{2}\kappa ''
+4a_{A}t^{2}(\dot{\phi}\dot{\kappa}-\phi '\kappa ')=0,
\end{eqnarray}
%%%%%%%%%%%%%%%%%%%%%%%%%%%%%%%%%%%%%%%%%%%%%%%%%%%%%%%%%%%%%%%%%%%%%%%%%%
\begin{eqnarray}
\label{ee-chi}
\ddot{\chi}-\chi''-(\dot{Z}+2a_{M}\dot{\phi})\dot{\chi}
+(Z'+2a_{M}\phi')\chi'
&+&(2X\dot{Z}-\dot{X})\dot{\omega}-(2XZ'-X')\omega' \nonumber \\
&+&e^{-2Z}X\{\dot{X}(X\dot{\omega}-\dot{\chi})+X'(X\omega'-\chi')\}
=0,
\end{eqnarray}
%%%%%%%%%%%%%%%%%%%%%%%%%%%%%%%%%%%%%%%%%%%%%%%%%%%%%%%%%%%%%%%%%%%%%%%%
\begin{eqnarray}
\label{ee-omega}
\ddot{\omega}-\omega''+(\dot{Z}-2a_{M}\dot{\phi})\dot{\omega}
-(Z'-2a_{M}\phi')\omega'+e^{-2Z}
\{\dot{X}(X\dot{\omega}-\dot{\chi})+X'(X\omega'-\chi')\}
=0,
\end{eqnarray}
%%%%%%%%%%%%%%%%%%%%%%%%%%%%%%%%%%%%%%%%%%%%%%%%%%%%%%%%%%%%%%%%%%%%%%%%%%%%
where $T^{ab}:=\frac{2}{\sqrt{-g}}\frac{\delta {\cal L}_{M}}{\delta g_{ab}}$ 
is the energy-momentum tensor.
Hereafter, we call the above system 
the {\it $T^{3}$-Gowdy symmetric EMDA} system. 

Note that the metric function $\lambda$ is decoupled with other functions.  
The function appears only in the Hamiltonian and momentum constraints
~(\ref{ce1}) and (\ref{ce2}).
Then, we can calculate the metric function $\lambda$ by evaluating the 
integral of $\lambda '$ from $-\pi$ to $-\pi$  after obtaining other functions 
from the evolution equations~(\ref{ee-z})-(\ref{ee-omega}).
%%%%%%%%%%%%%%%%%%%%%%%%%%%%%%%%%%%%%%%%%%%%%%%%%%%%%%%%%%%%%%%%%%%%%%%%%
%%%%%%%%%%%%%%%%%%%%%%%%%%%%%%%%%%%%%%%%%%%%%%%%%%%%%%%%%%%%%%%%%%%%%%%%%%%
\section{Global existence theorem in the areal coordinate}
\label{global}
%%%%%%%%%%%%%%%%%%%%%%%%%%%%%%%%%%%%%%%%%%%%%%%%%%%%%%%%%%%%%%%%%%%%%%%%%%%%
For simplicity, 
we will assume that the initial data are $C^{\infty}$ on $T^{3}$. 
In this circumstance we can show the following 
global existence theorem.
%%%%%%%%%%%%%%%%%%%%%%%%%%%%%%%%%%%%%%%%%%%%%%%%%%%%%%%%%%%%%%%%%%%%%%%%%%%%%%
\begin{theorem}
\label{main-Th}
Let $({\cal M}, g, \phi, \kappa, A)$ be the maximal globally hyperbolic 
development of the initial data for the Gowdy symmetric EMDA system.  
Then, ${\cal M}$ can be foliated by areal coordinate with $t\in (0,\infty)$.
\end{theorem}
%%%%%%%%%%%%%%%%%%%%%%%%%%%%%%%%%%%%%%%%%%%%%%%%%%%%%%%%%%%%%%%%%%%%%%%%%
{\it Proof of Theorem~\ref{main-Th}}:
The local existence and uniqueness of smooth solutions of 
the partial differential equation system~(\ref{ee-z})-(\ref{ee-omega}) 
follows from standard results for 
hyperbolic system~\cite{AS,CBG,FR,HL}.  
Then, it is sufficient to verify that for any globally hyperbolic subset of 
the $(t, \theta)$ cylinder on which they exist as a solution to
~(\ref{ee-z})-(\ref{ee-omega}), 
the functions $(Z, X, \phi, \kappa, \chi, \omega)$ and 
their first and second derivatives are uniformly bounded~\cite{MA}.  
To do this, we will use {\it light cone
estimate}~\cite{AH,BCIM,M81,M97}.

Let us now define the quadratic forms $\cal{G}$ and $\cal{H}$ by 
%%%%%%%%%%%%%%%%%%%%%%%%%%%%%%%%%%%%%%%%%%%%%%%%%%%%%%%%%%%%%%%%%%%%%%%
\begin{eqnarray}
\cal{G}&:=&
\frac{1}{2}t\left[\dot{Z}^{2}+Z'^{2}
+e^{-2Z}\left(\dot{X}^{2}+X'^{2}\right)\right] 
\nonumber \\
&& 
+2e^{-2a_{M}\phi}
\left[
e^{-Z}\{(X\dot{\omega}-\dot{\chi})^{2}+(X\omega'-\chi')^{2}\}
+e^{Z}(\dot{\omega}^{2}+\omega'^{2})\right]
\nonumber \\
&& 
+2t\left(\dot{\phi}^{2}+\phi '^{2}\right)
+\frac{1}{2}e^{4a_{A}\phi}t\left(\dot{\kappa}^{2}+\kappa'^{2}\right),
%-2\kappa(e^{-Z}(1-X^{2})-e^{Z})(\dot{\omega}\chi '-\omega '\dot{\chi}),
\end{eqnarray}
%%%%%%%%%%%%%%%%%%%%%%%%%%%%%%%%%%%%%%%%%%%%%%%%%%%%%%%%%%%%%%%%%%%%%%%%%%%
and 
\begin{eqnarray}
{\cal H} :=t\left[\dot{Z}Z'+e^{-2Z}\dot{X}X'\right]
+4e^{-2a_{M}\phi}\{e^{-Z}(X\dot{\omega}-\dot{\chi})(X\omega'-\chi')+
e^{Z}\dot{\omega}\omega '\}
+4t\dot{\phi}\phi '+te^{4a_{A}\phi}\dot{\kappa}\kappa'.
\end{eqnarray}
%%%%%%%%%%%%%%%%%%%%%%%%%%%%%%%%%%%%%%%%%%%%%%%%%%%%%%%%%%%%%%%%%%%%%%%%%%
Deriving $\cal{G}$ and $\cal{H}$ with respect to $t$ and $\theta$ and using 
the evolution equations~(\ref{ee-z})-(\ref{ee-omega}), 
after long calculation, 
we have the following inequalities.
%%%%%%%%%%%%%%%%%%%%%%%%%%%%%%%%%%%%%%%%%%%%%%%%%%%%%%%%%%%%%%%%%%%%%%%
\begin{eqnarray}
\label{g+h}
\sqrt{2}\partial_{\eta}(\cal{G}+\cal{H})&=&\frac{1}{2}\left[-\dot{Z}^{2}+Z'^{2}
+e^{-2Z}(-\dot{X}^{2}+X'^{2})\right]
+2\left[-\dot{\phi}^{2}+\phi'^{2}
+\frac{1}{4}e^{4a_{A}\phi}(-\dot{\kappa}^{2}+\kappa'^{2})\right] \nonumber \\
&:=&J\leq \frac{1}{t}\cal{G},
\end{eqnarray}
%%%%%%%%%%%%%%%%%%%%%%%%%%%%%%%%%%%%%%%%%%%%%%%%%%%%%%%%%%%%%%%%%%%%%%%%%%%%%
and 
%%%%%%%%%%%%%%%%%%%%%%%%%%%%%%%%%%%%%%%%%%%%%%%%%%%%%%%%%%%%%%%%%%%%%%%%%%%
\begin{eqnarray}
\label{g-h}
\sqrt{2}\partial_{\xi}(\cal{G}-\cal{H})&=&\frac{1}{2}\left[-\dot{Z}^{2}+Z'^{2}
+e^{-2Z}(-\dot{X}^{2}+X'^{2})\right]
+2\left[-\dot{\phi}^{2}+\phi'^{2}
+\frac{1}{4}e^{4a_{A}\phi}(-\dot{\kappa}^{2}+\kappa'^{2})\right] \nonumber \\
&:=&L\leq \frac{1}{t}\cal{G},
\end{eqnarray}
%%%%%%%%%%%%%%%%%%%%%%%%%%%%%%%%%%%%%%%%%%%%%%%%%%%%%%%%%%%%%%%%%%%%%%%%%
where 
$\partial_{\xi}:= \frac{1}{\sqrt{2}}(\partial_{t}+\partial_{\theta})$
and
$\partial_{\eta}:=\frac{1}{\sqrt{2}}(\partial_{t}-\partial_{\theta})$.  

At first, let us consider the future (expanding) direction.  
Integrating these equations (\ref{g+h}) and (\ref{g-h}) along null paths starting at 
$(\hat{\theta}, \hat{t})$ and ending at the initial $t_{0}$-surface 
(i.e. from the future into the past) and adding them we have
%%%%%%%%%%%%%%%%%%%%%%%%%%%%%%%%%%%%%%%%%%%%%%%%%%%%%%%%%%%%%%%%%%%%%%%%%%%
\begin{eqnarray}
\label{null-int}
{\cal G}(\hat{\theta}, \hat{t})&=&{\cal G}(\hat{\theta}+t_{0}-\hat{t}, t_{0})+
{\cal G}(\hat{\theta}-t_{0}+\hat{t}, t_{0}) \nonumber \\
&+&{\cal H}(\hat{\theta}+t_{0}-\hat{t}, t_{0})
-{\cal H}(\hat{\theta}-t_{0}+\hat{t}, t_{0}) \nonumber \\
&+&\int^{\hat{t}}_{t_{0}}[J(\hat{\theta}+s-\hat{t}, s)+
L(\hat{\theta}-s+\hat{t}, s)]ds.
\end{eqnarray}
%%%%%%%%%%%%%%%%%%%%%%%%%%%%%%%%%%%%%%%%%%%%%%%%%%%%%%%%%%%%%%%%%%%%%%%%%%%%
Next, we take supremums over all values of $\theta$ on the both sides of 
the equation~(\ref{null-int}).  
Then, we have 
%%%%%%%%%%%%%%%%%%%%%%%%%%%%%%%%%%%%%%%%%%%%%%%%%%%%%%%%%%%%%%%%%%%%%%%%%%%%
\begin{eqnarray}
\label{sup}
\sup_{\theta}{\cal G}(\theta, \hat{t})&\leq &
2\sup_{\theta}{\cal G}(\theta, t_{0})+
2\sup_{\theta}{\cal H}(\theta, t_{0}) \nonumber \\
&+&\int^{\hat{t}}_{t_{0}}\frac{1}{s}\sup_{\theta}{\cal G}(\theta, s)ds,
\end{eqnarray}
%%%%%%%%%%%%%%%%%%%%%%%%%%%%%%%%%%%%%%%%%%%%%%%%%%%%%%%%%%%%%%%%%%%%%%%%%
where we used the estimates (\ref{g+h}) and (\ref{g-h}). 
We can apply the following lemma to (\ref{sup}).
%%%%%%%%%%%%%%%%%%%%%%%%%%%%%%%%%%%%%%%%%%%%%%%%%%%%%%%%%%%%%%%%%%%%%%%%%%%
\begin{lemma}[Gronwall's lemma~\cite{HL}]
\label{GL}
Suppose that $\phi(t)$, $a(t)$, $b(t)\geq 0$.  
If 
$$
\phi(t)\leq a(t)+\int^{t}_{t_{0}}b(s)\phi(s)ds,
$$
then, 
$$
\phi(t)\leq a(t)+\int^{t}_{t_{0}}a(r)b(r)\exp{\int^{t}_{r}b(s)ds}dr.
$$
\hfill$\Box$
\end{lemma}
%%%%%%%%%%%%%%%%%%%%%%%%%%%%%%%%%%%%%%%%%%%%%%%%%%%%%%%%%%%%%%%%%%%%%%%%
By Lemma~\ref{GL}, we get the following inequality
\begin{eqnarray}
\label{G-ineq}
\sup_{\theta}{\cal G}(\theta, \hat{t})\leq 
2[\sup_{\theta}{\cal G}(\theta, t_{0})+
\sup_{\theta}{\cal H}(\theta, t_{0})]\exp(\int^{\hat{t}}_{t_{0}}\frac{1}{s}ds).
\end{eqnarray}
%%%%%%%%%%%%%%%%%%%%%%%%%%%%%%%%%%%%%%%%%%%%%%%%%%%%%%%%%%%%%%%%%%%%%%
From equation~(\ref{G-ineq}), since $\exp(\int^{\hat{t}}_{t_{0}}\frac{1}{s}ds)$ is bounded, 
we get the desired bound on 
$\mid\dot{Z}\mid$, $\mid Z'\mid$, $\mid e^{-Z}\dot{X}\mid$, 
$\mid e^{-Z}X'\mid$, 
$\mid e^{-a_{M}\phi-\frac{1}{2}Z}(X\dot{\omega}-\dot{\chi})\mid$, 
$\mid e^{-a_{M}\phi-\frac{1}{2}Z}(X\omega'-\chi')\mid$, 
$\mid e^{-a_{M}\phi+\frac{1}{2}Z}\dot{\omega}\mid$,
$\mid e^{-a_{M}\phi+\frac{1}{2}Z}\omega'\mid$, 
$\mid\dot{\phi}\mid$, $\mid \phi '\mid$, 
$\mid e^{2a_{A}\phi}\dot{\kappa}\mid$ and 
$\mid e^{2a_{A}\phi}\kappa '\mid$ for all $t\in (t_{0},\infty)$.

Once we have bounds on the first derivatives of $Z$ and $\phi$, 
it follows that $Z$ and $\phi$ are bounded for all 
$t\in (t_{0},\infty)$ as well since 
$$
u(t,x)=u(t_{0},x)+\int_{t_{0}}^{t}\partial_{s}u(s,x)ds, 
$$ 
where $u(t,x)$ is a function. 
Then, we have bounds on $\dot{X}$, $X'$, 
$X\dot{\omega}-\dot{\chi}$, $X\omega'-\chi'$, 
$\dot{\omega}$, $\omega'$, 
$\dot{\kappa}$ and $\kappa '$. 
Consequently, we obtain bounds on $X$, $\omega$ and $\kappa$ and 
furthermore, we have bounds on $\dot{\chi}$ and $\chi'$.  
Finally, we have bounds on $\chi$.  Thus, it has shown the boundedness 
of the zeroth and first derivatives of all the function except for 
$\lambda$.

Next, we must show bounds on the second derivatives of the functions.  
There is a well-known general fact that in order to ensure the continued 
existence of a solution of a system of semilinear wave equations 
it is enough to bound the first derivative pointwise 
(see Theorem 1.1 of Chapter III and page 42 of Ref.~\cite{AS}). 
Our system is a special case of that.  
Then, we have boundedness of the higher derivatives.

The same argument can be applied into the past (contracting) direction. 
By the constraint equations~(\ref{ce1}) and (\ref{ce2}), 
boundedness for the function $\lambda$ is 
also shown. 
Indeed, we have the following equations:
%%%%%%%%%%%%%%%%%%%%%%%%%%%%%%%%%%%%%%%%%%%%%%%%%%%%%%%%%%%%%
\begin{eqnarray}
\dot{\lambda}=P \hspace{1cm} {\rm and} \hspace{1cm}\lambda'=Q.
\end{eqnarray}
%%%%%%%%%%%%%%%%%%%%%%%%%%%%%%%%%%%%%%%
Since $P$ and $Q$ are bounded as we have already seen, 
$\dot{\lambda}$ and $\lambda'$ are 
bounded. 
Then, we have shown that the first derivatives of $\lambda$ 
must be bounded uniformly for all $0<t<\infty$. 
Consequently, $\lambda$ itself also is bounded. 
Concerning the second derivative of $\lambda$, 
we can apply the argument of Alinhac~\cite{AS} again. 
Then, we have uniform $C^{2}$ bounds on all of the functions for all 
$t\in (0,\infty)$.

Finally, we must demand that the function $\lambda$ is compatible with 
the periodicity in $\theta$. 
The following argument is similar with Moncrief's one~\cite{M81}. 
This is true if $\lambda(t,-\pi)=\lambda(t,\pi)$ 
over the interval of existence. 
Integrating equation~(\ref{ce2}) for the interval $\theta\in[-\pi,\pi]$, 
we have a constraint
%%%%%%%%%%%%%%%%%%%%%%%%%%%%%%%%%%%%%%%%%%%%%%%%%%%%%%%%%%%%%%%%
\begin{eqnarray}
\int^{\pi}_{-\pi}d\theta \{t[\dot{Z}Z'+e^{-2Z}\dot{X}X']-4tT_{t\theta}\}=0.
\end{eqnarray}
%%%%%%%%%%%%%%%%%%%%%%%%%%%%%%%%%%%%%%%%%%%%%%%%%%%%%%%%%%%%%%%%%
This constraint condition need only be imposed on the initial Cauchy surface 
since this integral is conservation on any time interval if all other 
functions satisfy the periodicity condition. 
This fact follows from the constraint equations~(\ref{ce1}) and (\ref{ce2}):
%%%%%%%%%%%%%%%%%%%%%%%%%%%%%%%%%%%%%%%%%%%%%%%%%%
\begin{eqnarray}
\frac{\partial}{\partial t}\{t[\dot{Z}Z'+e^{-2Z}\dot{X}X']-4tT_{t\theta}\}=
\frac{\partial}{\partial\theta}
\{\frac{t}{2}[\dot{Z}^{2}+Z'^{2}+e^{-2Z}(\dot{X}^{2}+X'^{2})]-2tT_{tt}\}.
\end{eqnarray}

Thus, we have completed the proof of Theorem~\ref{main-Th}.
\hfill$\Box$
%%%%%%%%%%%%%%%%%%%%%%%%%%%%%%%%%%%%%%%%%%%%%%%%%%%%%%%%%%%%%%%%%%%%%%%%%%%%%%%
\section{Existence of constant mean curvature foliations}
\label{CMC}
%%%%%%%%%%%%%%%%%%%%%%%%%%%%%%%%%%%%%%%%%%%%%%%%%%%%%%%%%%%%%%%%%%%%%%%%%%%%%%
For vacuum $T^{3}$-Gowdy spacetimes 
it has been proved that conjecture~\ref{Cj-ME} is valid~\cite{IM}.  
In that paper, the important idea is that one can apply the estimates 
for the functions in the areal time coordinate to show the existence of CMC 
foliations.  
Using the same argument with Ref.~\cite{IM}, 
we can show a global existence theorem in the CMC time coordinate 
for the $T^{3}$-Gowdy symmetric 
EMDA system.  
  
For the Gowdy symmetric spacetime 
the Einstein-matter equations imply that 
%%%%%%%%%%%%%%%%%%%%%%%%%%%%%%%%%%%%%%%%%%%%%%%%%%%%%%%%%%%%%%%
\begin{eqnarray}
-tr K(t)=e^{-\frac{\lambda}{4}}
t^{\frac{1}{4}}\left[\frac{1}{4}\dot{\lambda}
+\frac{3}{4t}\right]=e^{-\frac{\lambda}{4}}
t^{\frac{1}{4}}{\cal B}_{t},
\end{eqnarray}
%%%%%%%%%%%%%%%%%%%%%%%%%%%%%%%%%%%%%%%%%%%%%%%%%%%%%%%%%%%%
where $-tr K(t)$ is mean curvature of 
a compact Cauchy surface $S_{t}$ and 
%%%%%%%%%%%%%%%%%%%%%%%%%%%%%%%%%%%%%%%%%%%%%%%%%%%%%%%%%%%%%%%%%
\begin{eqnarray}
\label{bt}
{\cal B}_{t}:=\frac{3}{4t}+\frac{t}{4}\left[\dot{Z}^{2}+Z'^{2}
+e^{-2Z}(\dot{X}^{2}+X'^{2})\right]+tT_{tt}. 
\end{eqnarray}
%%%%%%%%%%%%%%%%%%%%%%%%%%%%%%%%%%%%%%%%%%%%%%%%%%%%%%%%%%%%
From the positivity (see equation~(\ref{ttt})) and bounds on the functions 
of right-hand-side 
of equation~(\ref{bt}), 
we have the following estimate.  
%%%%%%%%%%%%%%%%%%%%%%%%%%%%%%%%%%%%%%%%%%%%%%%%%%%%%
\begin{eqnarray}
\inf_{S_{t}}(-tr K(t))\geq Ct^{-\frac{3}{4}},
\end{eqnarray}
%%%%%%%%%%%%%%%%%%%%%%%%%%%%%%%%%%%%%%%%%%%%%%%%%%%%%%
for some constant $C$ and for all $0<t<t_{0}$.
This means that the spacetime has a crushing singularity as 
$t\rightarrow 0$ since the mean curvature goes to infinity 
uniformly~\cite{ME,M81}.

{\it Cosmological spacetimes} imply that spatially compact, 
globally hyperbolic spacetimes satisfying the strong energy condition
~\cite{BR}.  
Gerhardt has shown that in cosmological spacetimes 
if there exists a foliation whose mean curvature 
tends uniformly to infinity, there is a CMC foliation with same 
property~\cite{GC}.
Then, the past region of the initial hypersurface $S_{t_{0}}$, 
$D^{-}(S_{t_{0}})$, is covered by CMC hypersurfaces. 

Next, we must show that $D^{+}(S_{t_{0}})$ also can be covered by CMC 
hypersurfaces. 
To do this the following lemma is useful.  
Note that, from the previous argument,
we have one initial CMC hypersurface $\Sigma_{\tau_{0}}$
in $D^{-}(S_{t_{0}})$.
%%%%%%%%%%%%%%%%%%%%%%%%%%%%%%%%%%%%%%%%%%%%%%%%%%%%%%%%%%%%%%%%%%%%
\begin{lemma}
\label{CMC-lemma}
Suppose that 
$(D^{+}(\Sigma_{\tau_{0}}), g, \phi, \kappa, \chi)$ be a maximally extended, 
globally hyperbolic development of the initial data $\Sigma_{\tau_{0}}$ of 
$T^{3}$-Gowdy symmetric spacetime in the EMDA system.  
Then, 
$(D^{+}(\Sigma_{\tau_{0}}), g, \phi, \kappa, \chi)$ 
admit a unique, monotonic CMC foliation 
$i_{\tau}$ which covers $(D^{+}(\Sigma_{\tau_{0}}), g, \phi, \kappa, \chi)$.  
\end{lemma}
%%%%%%%%%%%%%%%%%%%%%%%%%%%%%%%%%%%%%%%%%%%%%%%%%%%%%%%%%%%%%%%%%%%%%%%%%%%%%%%%%%%
{\it Proof of Lemma~\ref{CMC-lemma}}:
The argument is very similar with one of the proof of Lemma 1 of Ref.~\cite{IM}.  
It is well known that there is a unique, monotonic, local CMC foliation of 
$(D^{+}(\Sigma_{\tau_{0}}), g, \phi, \kappa, \chi)$ defined near the 
initial hypersurface 
$\Sigma_{\tau_{0}}$~\cite{MT} and that the spacelike Killing fields must be 
tangent to CMC hypersurfaces~\cite{AL99}.  

First, we shall show that this foliation is uniformly spacelike.  
If $\Sigma_{\tau}=i_{\tau}(T^{3})$ is a CMC hypersurface in the local 
foliation, there is a smooth function $h_{\tau}:S^{1}\rightarrow R^{+}$ 
such that $\Sigma_{\tau}$ is defined in the coordinates $(t,\theta,y,z)$ 
by $t=h_{\tau}(\theta)\in [t_{0},t_{1}]$ 
and $h_{\tau}$ satisfies the following equation:
%%%%%%%%%%%%%%%%%%%%%%%%%%%%%%%%%%%%%%%%%%%%%%%%%%%%%%%%%%%%%%%%%%%%
\begin{eqnarray}
\label{embedded}
-\frac{d}{d\theta}\left[\frac{e^{\frac{\lambda}{4}}
t^{-\frac{1}{4}}h_{\tau}h_{\tau}'}{(1-(h_{\tau}'^{2}))^{1/2}}\right]=
e^{\frac{\lambda}{4}}
t^{-\frac{1}{4}}h_{\tau}
\left[(1-(h_{\tau}'^{2}))^{1/2}(\frac{1}{4}\dot{\lambda}
+\frac{3}{4t})+tr Ke^{\frac{\lambda}{4}}
t^{-\frac{1}{4}}\right]\Big|_{t=h_{\tau}(\theta)},
\end{eqnarray}
%%%%%%%%%%%%%%%%%%%%%%%%%%%%%%%%%%%%%%%%%%%%%%%%%%%%%%%%%%%%%%%%%%%%
where $h_{\tau}':=\frac{dh_{\tau}}{d\theta}$, 
$\lambda=\lambda(h_{\tau}(\theta),\theta)$ and $-tr K$ is the 
(constant) mean curvature of the embedded 
hypersurface. 
If $\mid h_{\tau}'\mid<1$, $\Sigma_{\tau}$ is spacelike 
since the induce metric $\gamma_{\tau}$ on $\Sigma_{\tau}$ is 
%%%%%%%%%%%%%%%%%%%%%%%%%%%%%%%%%%%%%%%%%%%%%%%%%%%%%%%%%%%%%%%%%%
\begin{eqnarray}
\gamma_{\tau}=e^{\frac{\lambda}{2}}t^{-\frac{1}{2}}
(1-(h_{\tau}'^{2}))d\theta ^{2}+
t\left[e^{-Z(t,\theta)}\left(dy+X(t,\theta)dz\right)^{2}
+e^{Z(t,\theta)}dz^{2}\right].
\end{eqnarray}
%%%%%%%%%%%%%%%%%%%%%%%%%%%%%%%%%%%%%%%%%%%%%%%%%%%%%%%%%%%%%%%%%%%%%
Integrating equation~(\ref{embedded}) 
from $\theta_{0}$ to $\theta_{1}$, we get 
%%%%%%%%%%%%%%%%%%%%%%%%%%%%%%%%%%%%%%%%%%%%%%%%%%%%%%%%%%%%%%%%%%%%
\begin{eqnarray}
\label{spacelikeCMC}
\Big|\frac{e^{\frac{\lambda}{4}}
t^{-\frac{1}{4}}h_{\tau}h_{\tau}'}{(1-(h_{\tau}'^{2}))^{1/2}}\Big|
\Big|_{\theta=\theta_{1}}\leq
\int^{2\pi}_{0}d\theta\left[
e^{\frac{\lambda}{4}}
t^{-\frac{1}{4}}h_{\tau}\left[(1-(h_{\tau}'^{2}))^{1/2}
{\cal B}_{t}
%(\frac{1}{4}\dot{\lambda}+\frac{3}{4t})
+\mid tr K\mid e^{\frac{\lambda}{4}}
t^{-\frac{1}{4}}\right]\right]\Big|_{t=h_{\tau}(\theta)},
\end{eqnarray}
%%%%%%%%%%%%%%%%%%%%%%%%%%%%%%%%%%%%%%%%%%%%%%%%%%%%%%%%%%%%%%%%%%
where $\theta_{1}$ is arbitrary and we can choose that 
$h_{\tau}'(\theta_{0})=0$.  
From the boundedness of the right-hand-side 
of equation~(\ref{spacelikeCMC}) and 
the fact that $h_{\tau}(\theta)\in[t_{0},t_{1}]\subset (0,\infty)$, 
we have that $\frac{h_{\tau}'}{(1-(h_{\tau}'^{2}))^{1/2}}$ is bounded 
by a constant 
and then, the estimate $\mid h_{\tau}'\mid<1$ is obtained.  
Thus, the CMC hypersurfaces are uniformly spacelike.  

The above argument holds for any $[t_{0},t_{1}]\subset (0,\infty)$. 
Then, the local foliation $i_{\tau}$ can be extended to $t\rightarrow \infty$. 
Next, we shall show that the leaves $\Sigma_{\tau}$ cannot approach 
the boundary 
at $t\rightarrow \infty$ without foliating a full neighborhood of 
the boundary.  
Since $\theta\in[0,2\pi)$ and $\mid h_{\tau}'\mid<1$, we have 
$$
0\leq \sup_{i_{\tau}(T^{3})}(t)-\inf_{i_{\tau}(T^{3})}(t)\leq 2\pi.
$$  
Then, the hypersurface $i_{\tau}(T^{3})$ cannot approach the boundary at 
$t\rightarrow \infty$ without foliating a full neighborhood of the boundary. 
(See also the proof of theorem 6.2 of~\cite{ARR}.)
\hfill$\Box$
%%%%%%%%%%%%%%%%%%%%%%%%%%%%%%%%%%%%%%%%%%%%%%%%%%%%%%%%%%%%%%%%%%%%%%%%%%

Lemma~\ref{CMC-lemma} states that future Cauchy development of the 
initial Cauchy surface is cover by CMC foliations.  
Combining the result from Gerhardt's theorem, which states that 
past Cauchy development of the 
initial Cauchy surface is cover by CMC foliations, 
we have the following theorem 
which supports the validity of conjecture~\ref{Cj-ME}. 
%%%%%%%%%%%%%%%%%%%%%%%%%%%%%%%%%%%%%%%%%%%%%%%%%%%%%%%%%%%%%%%%%%%%%%%%%%%%%%
\begin{theorem}
\label{main-Th2}
The spacetime can be covered by hypersurfaces of CMC, 
$-tr K(t)\in (0,\infty)$.  
\hfill$\Box$
\end{theorem}
%%%%%%%%%%%%%%%%%%%%%%%%%%%%%%%%%%%%%%%%%%%%%%%%%%%%%%%%%%%%%%%%%%%%%%%%%%%

%%%%%%%%%%%%%%%%%%%%%%%%%%%%%%%%%%%%%%%%%%%%%%%%%%%%%%%%%%%%%%%%%%%%%%%%%%%%%%
\section{Discussion}
\label{dis}
The remaining open question to prove the conjecture~\ref{Cj-AL} is to 
prove the inextendibility of the spacetime. 
This can be divided two questions which 
are: (1) Does the Kretschmann scalar of the spacetime blow up 
tend to singularities? 
(2) Is the spacetime geodesic complete into the future? 

For question (1), we have a very useful tool to analyze 
the nature of singularities, that is, the {\it Fuchsian 
algorithm} developed by Kichenassamy and Rendall~\cite{KR}. 
It has been shown that $T^{3}$-Gowdy spacetimes have an
asymptotically velocity-terms dominated (AVTD) singularity {\it in general 
in the sense that a family of solutions depends on the maximal number 
of arbitrary functions}.  
Recently this result has been generalized to the case of the Einstein-scalar 
system without symmetry assumptions~\cite{AR}, 
of the EMDA system 
with Gowdy symmetry~\cite{NTM} 
and of the $D$-dimensional Einstein-dilaton-p-form system 
without symmetry assumptions~\cite{DHRW}.  
These spacetimes are inextendible beyond the singularity since 
the Kretschmann scalar blows up there.  
We can conclude the inextendibility {\it in general} 
if we can answer the following question: 
Is there an open set of initial data 
on a regular Cauchy surface whose singularity is AVTD?
For the last question, we have only result 
for vacuum Gowdy spacetimes~\cite{CIM,RH}. 
We may extend the technique in their paper to our case.

Concerning question (2), 
Rein has obtained a related result for the Einstein-Vlasov system with 
hyperbolic symmetry~\cite{RG}. 
The spacetimes are shown to be causally geodesically complete in the 
future (expanding) direction if the data satisfy a certain size restriction.

If all of the above were proved, we would complete a proof of 
the strong cosmic censorship conjecture 
in the $T^{3}$-Gowdy symmetric EMDA system.

\section*{Acknowledgements}
I would like to thank Lars Andersson, H\"akan Andreasson, James Isenberg, 
Alan Rendall for useful discussion and Yoshio Tsutsumi for reading the 
manuscript and suggesting several improvement. 
I also want to thank anonymous referees for helpful comments.

%%%%%%%%%%%%%%%%%%%%%%%%%%%%%%%%%%%%%%%%%%%%%%%%%%%%%%%%%%%%%%%%%%%%%%%%

\end{document}